\documentclass[aip,jmp]{revtex4-1}

\usepackage{lineno,hyperref}
\modulolinenumbers[5]

\usepackage{hyperref}

\usepackage[all]{xy}

\usepackage{mathrsfs}
\usepackage{amssymb}
\usepackage{amsfonts}
\usepackage{latexsym}
\usepackage{amsmath}
\usepackage{xcolor}
\usepackage{wrapfig}
\usepackage{floatflt}

\usepackage{float}

\usepackage{graphicx}

\def\lb{\label}

\newcommand{\er}[1]{\textrm{(\ref{#1})}}

 \def\cM{{\mathcal M}}       \def\mM{{\mathscr M}}

\def\t{\tau}

    \def\N{{\mathbb N}}   

\def\no{\noindent}

\def\el2{\ell^{\,2}}             \def\1{1\!\!1}

\newcommand{\ca}{\begin{cases}}
\newcommand{\ac}{\end{cases}}
\newcommand{\ma}{\begin{pmatrix}}
\newcommand{\am}{\end{pmatrix}}
\def\eq{\begin{equation}}
\def\qe{\end{equation}}
\def\[{\begin{equation}}
\def\]{\end{equation}}

\begin{document}

\title[Inverted Thompson's lamp paradox]
{Programming infinite machines}

\author{Anton A. Kutsenko}
\affiliation{ Jacobs University, 28759 Bremen, Germany}
%\altaffiliation[Also at ]{Saint-Petersburg State University,
%Universitetskaya nab. 7/9, St. Petersburg, 199034, Russia.}

\date{\today}% It is always \today, today,

\begin{abstract}
For infinite machines which are free from the classical Thompson's
lamp paradox we show that they are not free from its inverted
version. We provide a program for infinite machines and an infinite
mechanism which simulate this paradox. While their finite analogs
work predictably, the program and the infinite mechanism demonstrate
an undefined behavior. As in the case of infinite Davies's machines,
our examples are free from infinite masses, infinite velocities,
infinite forces, etc. Only infinite divisibility of space and timeis assumed. Thus, the considered infinite devices are
possible in a continuous Newtonian Universe and they do not conflict
with continuous Newtonian mechanics. Some possible applications to
the analysis of the Navier-Stokes equations are discussed.
%The paper is a discussion note related to the article
%"Building infinite machines" by E. B. Davies published in
%British Journal for the Philosophy of Science.
\end{abstract}

%\pacs{47A10, 39A70, 78A40, 82D80}% PACS, the Physics and Astronomy
\keywords{infinite machines, hypercomputation, supertask, Thompson's
lamp paradox, beyond-Turing computers}
\maketitle

%\linenumbers
% 47A10, 39A70, 78A40, 82D80

\section{Introduction}\lb{S0}

The classical Thompson's lamp paradox appears in \cite{T1}. Let us provide its computer interpretation. Suppose that we have one byte $a$ of memory and some CPU which can carry out an infinite number of operations within a finite length of time. Consider the following set of instructions (so-called Zeno process)

$t=0:$\ \ \  $a:=0;$

$t=1/2:$\ \ \ $a:=1;$

$t=3/4:$\ \ \ $a:=0;$

$t=7/8:$\ \ \  $a:=1;$

....,

\no where $t$ is time. Assuming that CPU time of each next operation is twice faster than the CPU time of a previous operation, we can write a pascal code for the Thompson's program

$
 a:=0;
$

$
 \rm REPEAT
$

$
 a:= {\rm not}\ a;
$

$
 \rm UNTIL\ FALSE;
$

\no The paradox is that, we can not predict or determine the value of $a$ after the time $t$ when all operations are completed. For the first example, this time is $t=1$.

The theoretical description of infinite machines appears in \cite{BBJ1,H1}. The possibility of producing of such machines in certain exotic relativistic spacetimes (sometimes called Hogarth-Malament spacetimes) is demonstrated in \cite{EN1}. The construction of infinite machines in a continuous Newtonian Universe is discussed in \cite{D1}. In \cite{D1}, it is mentioned that the proposed infinite machine is free from the Thomson paradox. In this paper we show that such infinite machine is not free from the inverted Thompson's paradox. The rough idea of this paradox consists of changing the order of operations in the classical Thompson's paradox.

This topic is also closely related to the physical Church-Turing
thesis which is the conjecture that no computing device that is
physically realizable can exceed the computational barriers of a
Turing machine, see, e.g., \cite{W,ND,S}. The result of the paper
confirms this thesis since the infinite Davies's machine, which allows a hypercomputation, demonstrates also an unpredictable behaviour. This raises doubts about the fundamental possibility of constructing this machine and other hypercomputers (even without taking into account the quantum nature of the real world). Moreover, it indicates some fundamental 
difficulties in a continuous Newtonian Universe itself. In particular, this
observation may be helpful in analysis of Newtonian fluid dynamics,
e.g. in analysis of the Navier-Stokes equations. For example, if a
fluid analog of the mechanism considered in Section \ref{S3} exists
in a continuous Newtonian Universe then the Navier-Stokes equations
do not have a unique solution since the mechanism demonstrates an
undefined behavior. It is known that fluid motions can be very complex, see, e.g. \cite{FS}. They can create arbitrary small eddies and turbulent vortices with bizarre shapes. All this allows us to hope for the possibility of  constructing the fluid analogs that will be close in some properties to the mechanism depicted on Fig. \ref{fig4}. Then, it can be perspective for presenting a negative answer to the "millenium problem".

Another useful information about the physical Church-Turing thesis
along with hypercomputation and supertask can be found in
\cite{CR,C1,HK,C2,N1}. It is also useful to note the paradox called "paradox of predictability" or "second oracle paradox", see, e.g. \cite{RC}. The infinite analog of the paradox of predictability has some similar features with the inverted Thompson paradox. The corresponding analysis will be presented elsewhere.

The paper is organized as follows. Sections \ref{S1},\ref{S2}
contain the description of the infinite machine and the program
"puzzle" which demonstrates unpredictable behavior. Section \ref{S3}
contains the description of a pure mechanical device which
demonstrates the same undefined behavior as the program "puzzle". We
conclude in Section \ref{S4}.

\section{The construction\lb{S1}}

We consider a simplified version of infinite machine from \cite{D1}. The machine $\mM=\cup_{n=1}^{\infty}\cM_n$ consists of infinite number of finite machines $\cM_n$, $n\in\N$, see Fig. \ref{fig1}. The machine $\cM_{n+1}$ is a small copy of the machine $\cM_n$ for all $n$. The machine $\cM_{n+1}$ is also twice faster than the machine $\cM_n$ for all $n$. For instance, we assume that CPU time $\t_n$ of $\cM_n$ is equal to $1/2^n$ for all $n$. We do not assume that the memory of $\cM_{n+1}$ is large than the memory of $\cM_{n}$. All machines have the same memory size, say $1$ byte for data and $1$ Kbyte for a program code and for built-in variables. Single-threaded CPU (interpreter) of each $\cM_n$ can perform integer and logic operations and simple data manipulations. Each $\cM_n$ can interact directly with adjacent $\cM_{n+1}$ only.

\begin{figure}[H]
    \includegraphics[width=1\textwidth]{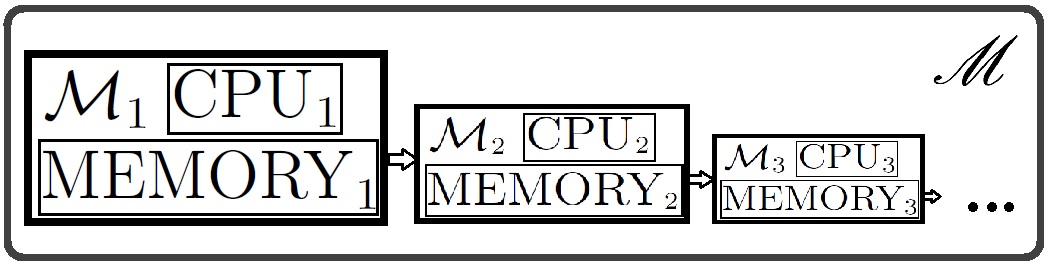}
    \caption{The structure of the infinite machine $\mM$.} \lb{fig1}
\end{figure}

Let us describe some commands of the machine $\mM$. If CPU of the machine $\cM_n$ gets the instruction
$$
 {\rm COPY\_PROGRAM\_NEXT}\ something;
$$
then it copies the code placed between "${\rm PROGRAM}\ something:$" and "${\rm END}\ something;$" to the program memory of $\cM_{n+1}$ and runs the copy there. The instruction
$$
 {\rm IDLE}\ m;
$$
says that CPU should skip $m$ CPU time's $\t_n$ before executing next instructions. CPU time $\t_n(=1/2^n)$ depends on the machine $\cM_n$, where the instruction ${\rm IDLE}$ is performed. Any $\cM_n$ has built-in variables:
$$
 {\rm VALUE}
$$
which refers to the byte data memory of $\cM_n$, and
$$
 {\rm VALUE\_NEXT}
$$
which refers to the byte data memory of $\cM_{n+1}$. At the
beginning of a program, all values are initialized to $0$. This
remark is very important in the program "puzzle" considered below.
The instructions
$$
 {\rm NOT},\ \ \ :=
$$
mean the bitwise "not" and the assignment operation respectively. In particular, $({\rm NOT}\ 1)=0$ and $({\rm NOT}\ 0)=1$. We assume that all instructions described above (except ${\rm IDLE}$) take one CPU time $\t_n$ for performing. The CPU time $\t_n(=1/2^n)$ depends on the machine $\cM_n$, where the instruction is performed.

The machine $\mM$ is free from the classical Thompson's paradox because the CPU-s can not manipulate with the fixed memory cell an infinite number of times. Nevertheless, $\mM$ is not free from the inverted Thompson's paradox.

\section{The puzzle\lb{S2}}
The following program emulates the inverted Thompson's paradox. The
code is written in a pascal-based programming language. The comments
are placed in parentheses $\{...\}$.
$$
 \begin{array}{ll}
 {\rm PROGRAM}\ puzzle: & \{entry\ point\} \\
 {\rm COPY\_PROGRAM\_NEXT}\ puzzle; & \{instruction\ 1\} \\
 {\rm IDLE}\ 2; & \{instruction\ 2\} \\
 {\rm VALUE}:={\rm NOT}\ {\rm VALUE\_NEXT}; & \{instruction\ 3\ (two\ instructions)\} \\
 {\rm END}\ puzzle; & \{exit\}
 \end{array}
$$
The program starts on $\cM_1$, copies to $\cM_2$ and starts there, waits for some time, takes the "inverted" value from $\cM_2$ and stops.
The same happens in $\cM_2$, $\cM_3$, and so on. In fact, the program works like a shader for a multiprocessor system.  The corresponding time diagram is plotted in Fig. \ref{fig2}. Let us denote the time when $i$-th instruction ($i=1,2,3$) starts on $\cM_n$ ($n\in\N$) by $t_{ni}$. Let us denote the exit time on $\cM_n$ by $t_{n4}$, $n\in\N$. Then
\begin{multline}\lb{001}
 t_{n1}=\sum_{m=1}^{n-1}\t_m=1-2^{1-n},\ \ \ t_{n2}=t_{n1}+\t_n=1-2^{-n},
\\
 t_{n3}=t_{n2}+2\t_n=1+2^{-n},\ \ \ t_{n4}=t_{n3}+2\t_n=1+2^{-n}+2^{1-   n}.
\end{multline}
Due to \er{001}, all values will be initialized and there are no conflicts between parallel programs working on different $\cM_n$, since only adjacent machines can interact. Nevertheless, we can not determine the value on $\cM_1$ at the end of the program "puzzle". The reason is similar as in the inverted Thomson paradox. Both values ${\rm VALUE}_1=0$ and ${\rm VALUE}_1=1$ are possible (and impossible) at the end of the program. More precisely, if we execute "puzzle" on the finite machine $\mM_N=\cup_{n=1}^N\cM_n$ (the cascade stops in $\cM_N$) then ${\rm VALUE}_1=1$ for even $N$ and ${\rm VALUE}_1=0$ for odd $N$. But for $N=\infty$ we can not say: is $N$ even or odd?
\begin{figure}[H]\centering
    \includegraphics[width=0.7\textwidth]{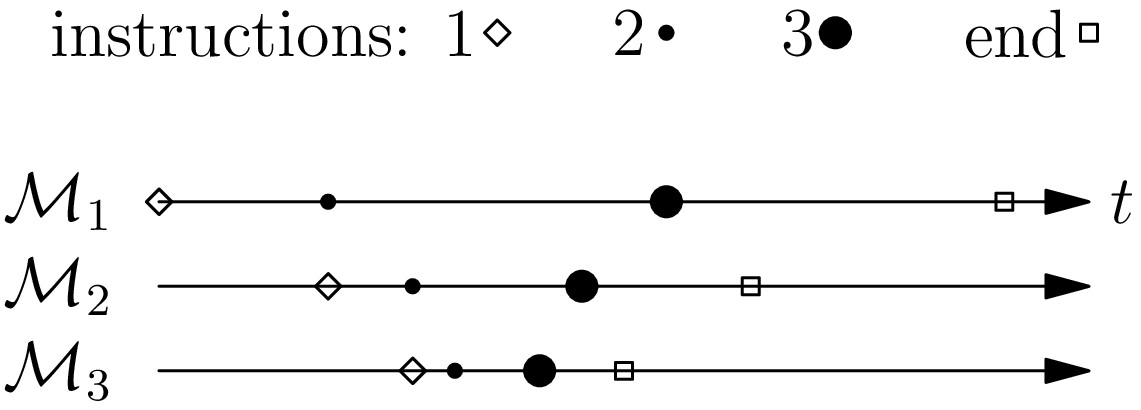}
    \caption{Starting time for each of instructions in machines $\cM_1$, $\cM_2$, $\cM_3$.} \lb{fig2}
\end{figure}

\section{The mechanical interpretation\lb{S3}}

Let us consider another variant of the inverted Thompson's lamp
paradox. Consider the mechanism "mousetrap" depicted in Fig.
\ref{fig3}. The mechanism consists of the beam on the spring. The
beam in tension (vertical position) is fixed with a thread. When
ball is tearing the tread, the beam latches horizontally and it does
not let through another ball.

\begin{figure}[H]\centering
    \includegraphics[width=0.99\textwidth]{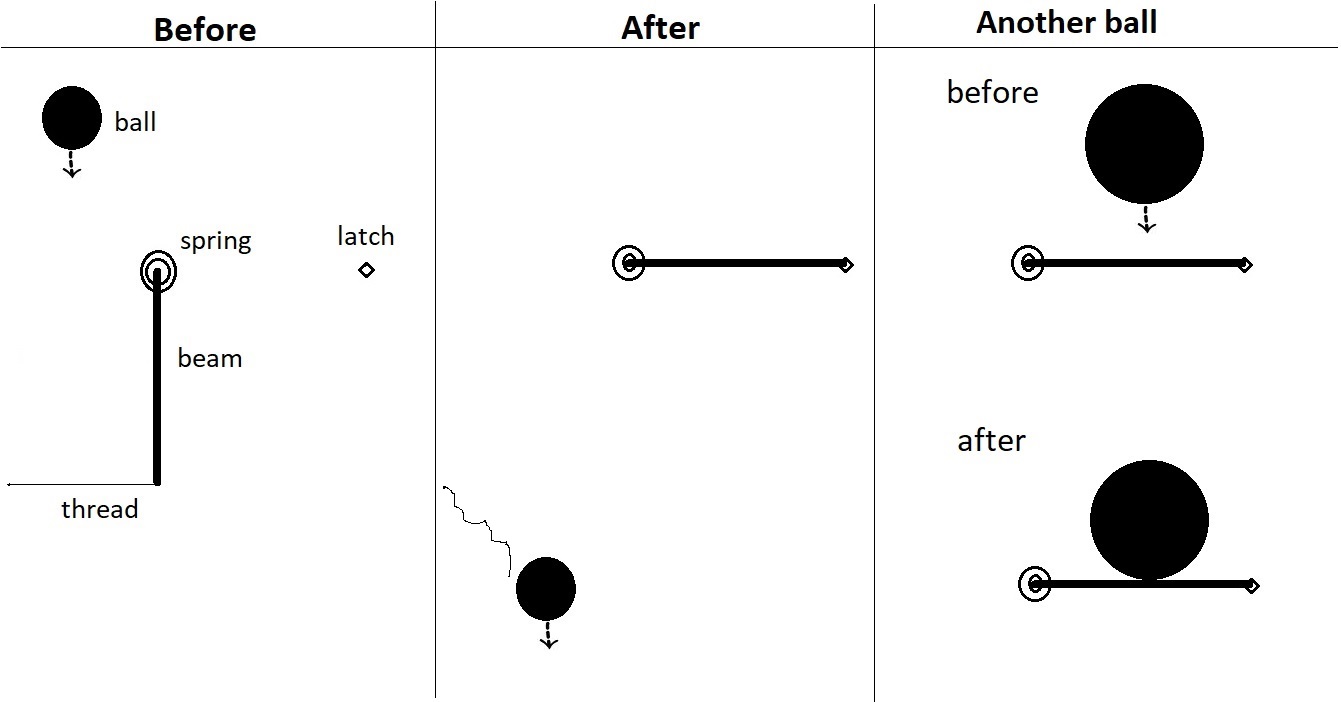}
    \caption{The structure of the finite mechanism.} \lb{fig3}
\end{figure}

Consider infinite number of the finite mechanisms depicted in Fig.
\ref{fig4}. Each next finite mechanism is a small (half of the size)
replica of the previous finite mechanism. To avoid various (e.g. centrifugal) effects, we can tune material properties of the spring and the beam of the next mechanism. We suppose also that there
are infinite number of balls that move with a same constant velocity
to the threads of the finite mechanisms. The size of each next ball
is twice smaller than the size of the previous ball. The distances
between the balls and the corresponding threads are chosen such that
the smaller ball can tear the thread before the larger ball can
reach the clipped horizontal beam corresponding to the smaller ball.
Thus, the larger ball can not tear its thread since the beam is
latched.

Note that any fixed constant value can be added to the distances
between the balls and the threads. It is useful if we want that the
smallest (limit) distance between the balls and the threads or beams
is non-zero. Thus, there is a non-zero time interval between the
start and the time when the balls reach the threads or latched
beams.

The behavior of the infinite mechanism from Fig. \ref{fig4} is
indeterminate. We can not predict: will the largest beam be in a
vertical or horizontal (latched) position after the balls fall down?
The reason is the same as in the programm "puzzle". If the number of
balls is a finite number, say $N$ then the largest beam is in a
horizontal position for odd $N$ and in a vertical position for even
$N$. But we can not say $N=\infty$ is even or odd number. Note that
in our example we do not assume infinite masses, velocities,
densities. So, the unpredictable infinite mechanism  may well exist
in a Newtonian Universe. Of course, such mechanism is not possible
in our world because of the principles of quantum mechanics.

\begin{figure}[H]\centering
    \includegraphics[width=0.99\textwidth]{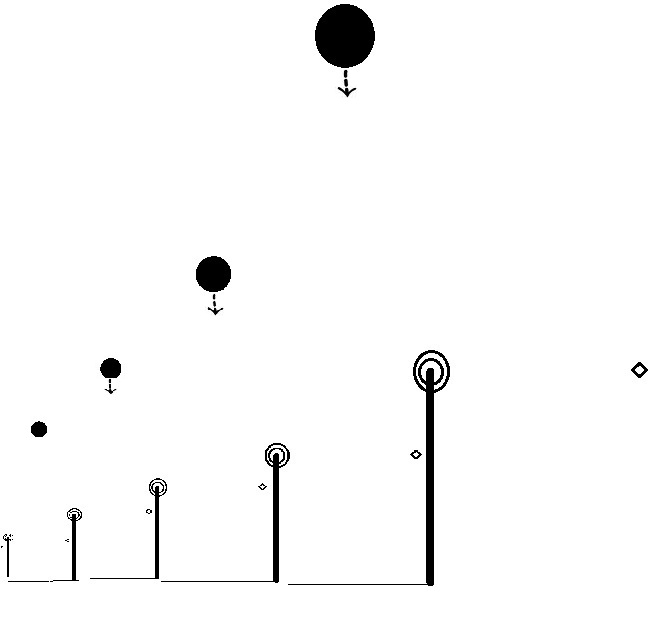}
    \caption{The infinite mechanism.} \lb{fig4}
\end{figure}

\section{Conclusion\lb{S4}}

Perhaps, any machine which uses the actual infinity is not free from
Thompson-type paradoxes. Even physically reasonable assumptions may
not be helpful. Probably, the main problem lies in our understanding
of infinity. Nevertheless, a part of our mind can successfully
develop infinite theories such as Peano arithmetic. Hence, there is
a natural question which, however, can not be formulated rigorously:
Is that part of our mind is an infinite machine and how it works?

%\section*{Acknowledgements}
%This work was partially supported by the RSF project
%N\textsuperscript{\underline{o}}15-11-30007 and DFG project TRR 181.

\bibliography{infinite}

\begin{thebibliography}{10}
\expandafter\ifx\csname url\endcsname\relax
  \def\url#1{\texttt{#1}}\fi
\expandafter\ifx\csname urlprefix\endcsname\relax\def\urlprefix{URL }\fi
\expandafter\ifx\csname href\endcsname\relax
  \def\href#1#2{#2} \def\path#1{#1}\fi

\bibitem{T1}
{J. Thompson}, {Tasks and supertasks}, Analysis 15 (1954) 1--13.

\bibitem{BBJ1}
G.~S. Boolos, J.~P. Burgess, R.~C. Jeffrey, Computability and logic, Fifth
  edition, Princeton University, United States, 2007.

\bibitem{H1}
M.~Hogarth, Predictability, computability, and spacetime, Cambridge University,
  1996, phd thesis.

\bibitem{EN1}
J.~Earman, J.~D. Norton, Infinite pains: The trouble with supertasks, in:
  A.~Morton, S.~P. Stich (Eds.), Benacerraf and his critics, Cambridge, MA:
  Blackwell, 1996, pp. 231--261.

\bibitem{D1}
{E. B. Davies}, {Building infinite machines}, Brit. J. Phil. Sci. 52 (2001)
  671--682.

\bibitem{W}
{C. Wuthrich}, {A quantum-information-theoretic complement to a
  general-relativistic implementation of a beyond-Turing computer}, Synthese
  192 (2015) 1989--2008.

\bibitem{ND}
{I. Nemeti, G. David}, {Relativistic computers and the Turing barrier}, Appl.
  Math. Comput. 178 (2006) 118--142.

\bibitem{S}
{M. Stannett}, {Computing the appearance of physical reality}, Appl. Math.
  Comput. 219 (2012) 54--62.

\bibitem{FS}
{G. Falkovich, K. R. Sreenivasan}, {Lessons from hydrodynamic turbulence},
  Physics Today 59 (2006) 43--49.

\bibitem{CR}
{P. Clark, S. Read}, {Hypertasks}, Synthese 61 (1984) 387--390.

\bibitem{C1}
{B. J. Copeland}, {Hypercomputation}, Mind. Mach. 12 (2002) 461--502.

\bibitem{HK}
{A. Hagar, A. Korolev}, {Quantum hypercomputability?}, Mind. Mach. 16 (2006)
  87--93.

\bibitem{C2}
{P. Cotogno}, {A brief critique of pure hypercomputation}, Mind. Mach. 19
  (2009) 391--405.

\bibitem{N1}
{A. Nayebi}, {Practical intractability: A critique of the hypercomputation
  movement}, Mind. Mach. 24 (2014) 275--305.

\bibitem{RC}
{S. Rummens, S. E. Cuypers}, {Determinism and the paradox of predictability},
  Erkenntnis 72 (2010) 233--249.

\end{thebibliography}

\end{document}